\documentclass[useAMS]{gGAF2e}
\newcommand{\lta}{\;
  \raise0.3ex\hbox{$<$\kern-0.75em\raise-1.1ex\hbox{$\sim$
  }}\;\hskip-2pt }
\newcommand{\gta}{\;
  \raise0.3ex\hbox{$>$\kern-0.75em\raise-1.1ex\hbox{$\sim$
  }}\;\hskip-2pt }
\newcommand{\apropto}{\;
  \raise0.3ex\hbox{$\propto$\kern-0.75em\raise-1.1ex\hbox{$\sim$
  }}\;\hskip-2pt }
\usepackage{graphicx}

\renewcommand{\Omega}{{\varOmega}}

\begin{document}
\doi{10.1080/03091920xxxxxxxxx}
 \issn{1029-0419} \issnp{0309-1929} \jvol{00} \jnum{00} \jyear{2009} 

\markboth{David Moss and Dmitry Sokoloff}{Magnetic field reversals
and galactic dynamos}

\title{Magnetic field reversals and galactic dynamos}

\author{DAVID MOSS${\dag}$ and DMITRY SOKOLOFF$^\ast\ddag$\thanks{$^\ast$Corresponding author. Email: d$_-$sokoloff@hotmail.com}
\vspace{6pt}\\\vspace{6pt}  ${\dag}$School of Mathematics,
University of Manchester, Oxford Road, Manchester, M13 9PL, UK\\
${\ddag}$ Department of Physics, Moscow State University, Moscow,
119992, Russia
\\\vspace{6pt}\received{Received 8 January 2012; in final form 23 May 2012; first published online ????}}

\maketitle

\begin{abstract}
We argue that global magnetic field reversals similar to those
observed in the Milky Way occur quite frequently in mean-field
galactic dynamo models that have relatively strong, random, seed
magnetic fields that are localized in discrete regions. The number
of reversals decreases to zero with reduction of the seed strength,
efficiency of the galactic dynamo and size of the spots of the seed
field. A systematic observational search for magnetic field
reversals in a representative sample of spiral galaxies promises to
give valuable information concerning seed magnetic fields and, in
this way, to clarify the initial stages of galactic magnetic field
evolution.
\bigskip


\end{abstract}

\section{Introduction}
\label{intro}
An outstanding question in the study of magnetic fields in spiral galaxies is
that of large-scale field reversals. Do they occur, how easily can they
be detected, can dynamo models explain them?

It appears that there may be such a field reversal in the Milky Way
-- certainly there is a localized feature relatively close to the
Sun, but the difficulties of observing through the confusion of the
disc plane make it difficult to be absolutely certain whether it is
global in extent. Further reversals have been claimed to exist, but
with increasing uncertainty  \citep[e.g.][]{sn79,b96,f01,menetal08,notkat10,kron11, vaneck11, beck11}.
Speaking generally, the idea that magnetic field reversals are
present in the Milky
Way is widely accepted.

The situation in external galaxies \cite[see the review][]{b96},
where the problem of observing from a position in the disc plane is
absent, is nevertheless rather more uncertain. Except for the
immediate neighbours of the Milky Way, the available resolution is
too low to make definitive statements. But in the well observed
nearby M31, reversals appear to be absent.

Classical mean-field dynamo models that start from a dynamically
small field do not in general produce field reversals \citep[e.g.][]{r88,m90},
because the leading eigenfunction of the kinematic
mean-field galactic dynamo usually has no reversals, and this
dominates the field structure as the field strength grows.

The situation becomes less straightforward when the seed field for
the dynamo is stronger in comparison to the contemporary
galactic magnetic field, so the galactic dynamo becomes nonlinear before
the magnetic field configuration approaches the leading
eigenfunction of the kinematic dynamo. It appears that
long-lived reversals can survive in the form of nonlinear fronts 
\citep{poezd,becketal94,metal98}: 
these are essentially field discontinuities, 
\citep[see e.g.][]{betal94,vetal94,mps00,psm01}.
However these models depend on
rather carefully chosen initial conditions, and existing examples
are one dimensional.

More recently, \cite{metal12} have published an extended mean-field
model in which long-lived reversals appear to be a quite common
feature. This dynamo model differs in several important ways from
earlier 2D mean-field models. Notably these computations begin from
initial conditions of small-scale, approximately equipartition
strength, fields that are randomly distributed in many discrete
spots. This is in contrast to nearly all previous mean-field models,
which begin from dynamically weak seed fields, either
quasi-homogeneous or small-scale. The rationale is that these
dynamically strong fields
are the result of small-scale dynamo action in star forming regions
in the protogalaxy. In the computations described in \cite{metal12}
there are also ongoing injections of small-scale field, but we will
argue here that these are not central to the generation of
reversals.

Our approach is close to that of the one dimensional model
of \cite{becketal94}, where
a localized random initial condition was considered and
long-lived large-scale magnetic
reversals developed. (This contrasts with the model of \cite{poezd}, where there
were ongoing injections of random fields.)
Thus both \cite{poezd} and \cite{becketal94} considered magnetic fields depending
on galactocentric radius only, i.e. ring-like magnetic structures were
described. The novelty of our approach is that we consider seed fields
that depend on both radius and azimuth, i.e. there is no {\it a priori} restriction
to ring-like structures, and any such structures develop as a result of the
 dynamo process. We demonstrate below that ring-like structures with reversals,
 similar to those described by \cite{becketal94}, also develop in our simulations.
Thus our model reinforces and extends the results of \cite{becketal94}.

On the other hand, we note that the stable magnetic ring-like structures
with reversals obtained by \cite{poezd} appeared in a model with
some fine tuning of the model parameters. Stability conditions for such
structures obtained by \cite{betal94} depend on rather subtle properties
of the galactic disc; this gave a hint that the magnetic configuration of
the Milky Way with field reversals is a rare exception
to the typical situation without reversals. Possibly, there was a
misinterpretation of the results, and maybe
the authors of the papers cited above were
not insistent enough.  Nevertheless, dynamo generated configurations with
reversals remained rather neglected by the astronomical community.
Moreover, the option of including continuous injection of small-scale
magnetic field as considered by \cite{poezd}, was not further developed
until recently.

In contrast, the new generation of dynamo models presented in \cite{metal12}
produce magnetic structures with reversals even for the simplest and
most primitive distribution of the dynamo governing parameters and they
appear to occur more or less as commonly as structures without reversals.
This is why we consider below examples of dynamo models with (and without)
reversals. They are demonstrably oversimplified and do not include any fine
tuning of the governing parameters of the dynamo.

The paper \cite{metal12} focussed on applications of the model to
future observations of magnetic fields of the earliest galaxies using the
recently developed (e.g. LOFAR) and planned (SKA) radio telescopes. The
aim of this paper is to describe the long-lived reversals obtained
as a quite general phenomenon of galactic dynamos. We are motivated here
by future observations of magnetic fields in very young galaxies,
whose detailed structures
 are almost unknown at the moment, and so we study a very simple, generic model.
This is in some contrast to the approch of \cite{poezd} (and to some extent \cite{becketal94}) who focussed their attention on the dynamo properties of
particular nearby galaxies (M31 and the Milky Way) and exploited particular
subtle properties of their rotation curves and other parameters.

\section{The dynamo model}

Detailed modeling of galactic magnetic of galactic magnetic field
evolution from the time of galactic formation up to the age of a
contemporary galaxy is a severe problem, both because of obvious numerical
difficulties and our very poor knowledge of
hydrodynamics of early galaxies. \cite{metal12} used a reasonable
simplification of the problem in the mean-field approach in the form of
the "no-$z$" model \citep[e.g.][]{m95}
which restricts the modelling to quantities which are accessible observationally,
at least in the immediate future, and make the
numerical implementation affordable. \cite{phil01} systematically
compared this approximation with the standard "local model" \citep[cf.][]{r88} 
and found satisfactory agreement.

Here we use the same simplified model.
For the sake of clarity we briefly reproduce the relevant
equations from \cite{metal12}.
The code solves in the $\alpha\omega$ approximation explicitly for
the field components parallel to the disc plane while the component
perpendicular to this plane (i.e. in the $z$-direction) is given by
the solenoidality condition. An even (quadrupole-like) magnetic
field parity with respect to the disc plane is assumed. The field
components parallel to the plane can be considered as mid-plane values,
or as a form of vertical average through the disc. The key
parameters are the aspect ratio $\lambda=h/R$, where $h$ corresponds
to the semi-thickness of the warm gas disc and $R$ is its radius,
and the dynamo numbers $R_\alpha=\alpha_0 h/\eta, R_\omega=\Omega_0
h^2/\eta$ (if disc thickness varies with radius then $h$ is some reference
value, $h_0$ say).  $\lambda$ must be a small parameter. $\eta$ is the
turbulent diffusivity, assumed uniform, and $\alpha_0, \Omega_0$ are
typical values of  the $\alpha$-coefficient and angular velocity
respectively. Thus the dynamo equations become  in cylindrical polar
coordinates $(r, \phi, z)$

\begin{equation}
\frac{\partial B_r}{\partial t} = -R_\alpha B_\phi-\frac{\pi^2}{4}
B_r +\lambda^2\left[\frac{\partial}{\partial
r}\left(\frac{1}{r}\frac{\partial}{\partial
r}(rB_r)\right)+\frac{1}{r^2}\frac{\partial^2B_r}{\partial\phi^2}-\frac{2}{r^2}\frac{\partial
B_\phi}{\partial\phi}\right] , \label{evolBr}
\end{equation}
\begin{equation}
\frac{\partial B_\phi}{\partial t} = R_\omega r
B_r\frac{{\mathrm d}\Omega}{{\mathrm d}r}-R_\omega\Omega\frac{\partial B_\phi}{\partial
\phi}-\frac{\pi^2}{4} B_\phi
+\lambda^2\left[\frac{\partial}{\partial
r}\left(\frac{1}{r}\frac{\partial}{\partial r}(rB_\phi)\right)
+\frac{1}{r^2}\frac{\partial^2B_\phi}{\partial
\phi^2}-\frac{2}{r^2}\frac{\partial B_r}{\partial \phi}\right],
\label{evolBphi}
\end{equation}
 where $z$ does not appear explicitly. Here the disc semi-thickness
$h$ is assumed to be constant: taking $h=h(r)$ would introduce
geometric correction factors $(h(r)/h_0), (h(r)/h_0)^2$ multiplying
$R_\alpha$, and the terms $-\frac{\pi^2}{4}(B_r, B_\phi)$, respectively.
This equation has been calibrated by introduction of the factors
$\pi^2/4$ in the vertical diffusion terms. In principle in the
$\alpha\omega$  approximation the parameters $R_\alpha, R_\omega$
can be combined into a single dynamo number $D=R_\alpha R_\omega$,
but we choose to keep them separate. Equations (1) and (2) are displayed
in polar coordinates but for ease of numerical implementation they
are rewritten in Cartesians.
Computations are performed on a uniform grid of resolution
$229\times 229$ in the region $-1\le x,y\le 1$. This resolution was tested in \cite{metal12},
and indeed is in excess of what is required for the very straightforward dynamo
problem solved after $t=0$.
The unit of time is about $0.8$\,Gyr. Further details are given in \cite{metal12}.

Length, time and magnetic field are non-dimensionalized in units of
$R$, $h^2/\eta$ and the equipartition field strength $B_{\rm eq}$
respectively. A naive algebraic $\alpha$-quenching nonlinearity is
assumed, $\alpha=\alpha_0/(1+B^2/B_{\rm eq}^2)$, where $B_{\rm eq}$
is the strength of the equipartition field in the general disc
environment.
Note that small-scale field is injected only at time zero; after that we conduct a standard mean field dynamo simulation.

The coefficients $\alpha_0$ and $B_{\rm eq}$
are  assumed to be uniform in the work
described in this paper
because our restricted knowledge of the properties of the earliest
galaxies give no secure basis for
more realistic assumptions. For example, it is argued that $\alpha_0$ depends
on both disc thickness and local angular velocity,  i.e.
$\alpha_0\propto \Omega h^{-1}$, where $h$ can be expected to increase with radius,
and $\Omega$ to decrease.
Of course, we continue our simulations to a time corresponding to the present
day where knowledge of particular galaxies is rather better -- we emphasize again that we are studying generic properties
of thin disc dynamos.  Taking typical galactic values, we can
estimate $R_\alpha={\mathrm O}(1)$, $R_\omega={\mathrm O}(10)$ (i.e. $D ={\mathrm O}(10)$).

We appreciate that more sophisticated approaches to the saturation
process exist, however it does appear that in some cases at least,
such a naive algebraic quenching can reproduce reasonably well the
results from a  more sophisticated treatment \citep[e.g.][]{kletal02}.

\section{Results}
\label{res}

\cite{metal12} presented a particular galactic dynamo model
that quite often resulted in magnetic field configuration with one or
more reversals. We argue here that many specific features of the
model \cite{metal12} are not essential for the development of reversals.

\begin{figure}
\includegraphics[width=0.45\textwidth]{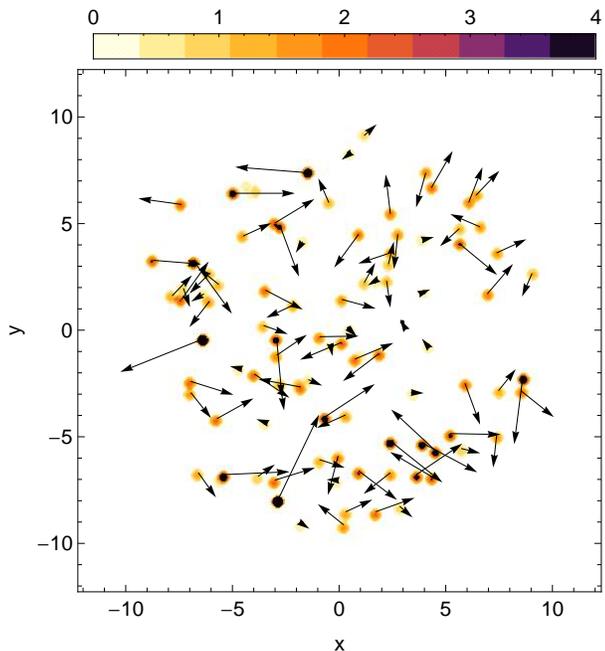}
\caption{(Colour online).  A typical configuration of the initial magnetic field;
only the ${\bm B}$ vectors at the centre of the spots are shown.
The shaded circles indicate the gaussian half-width of the spots.
(Taken from \protect\cite{metal12}.) }
\label{initfield}
\end{figure}

\begin{figure*}
\begin{center}
\vspace{0.5cm}
\begin{tabular}{ll}
(a)\includegraphics[width=0.44\textwidth]{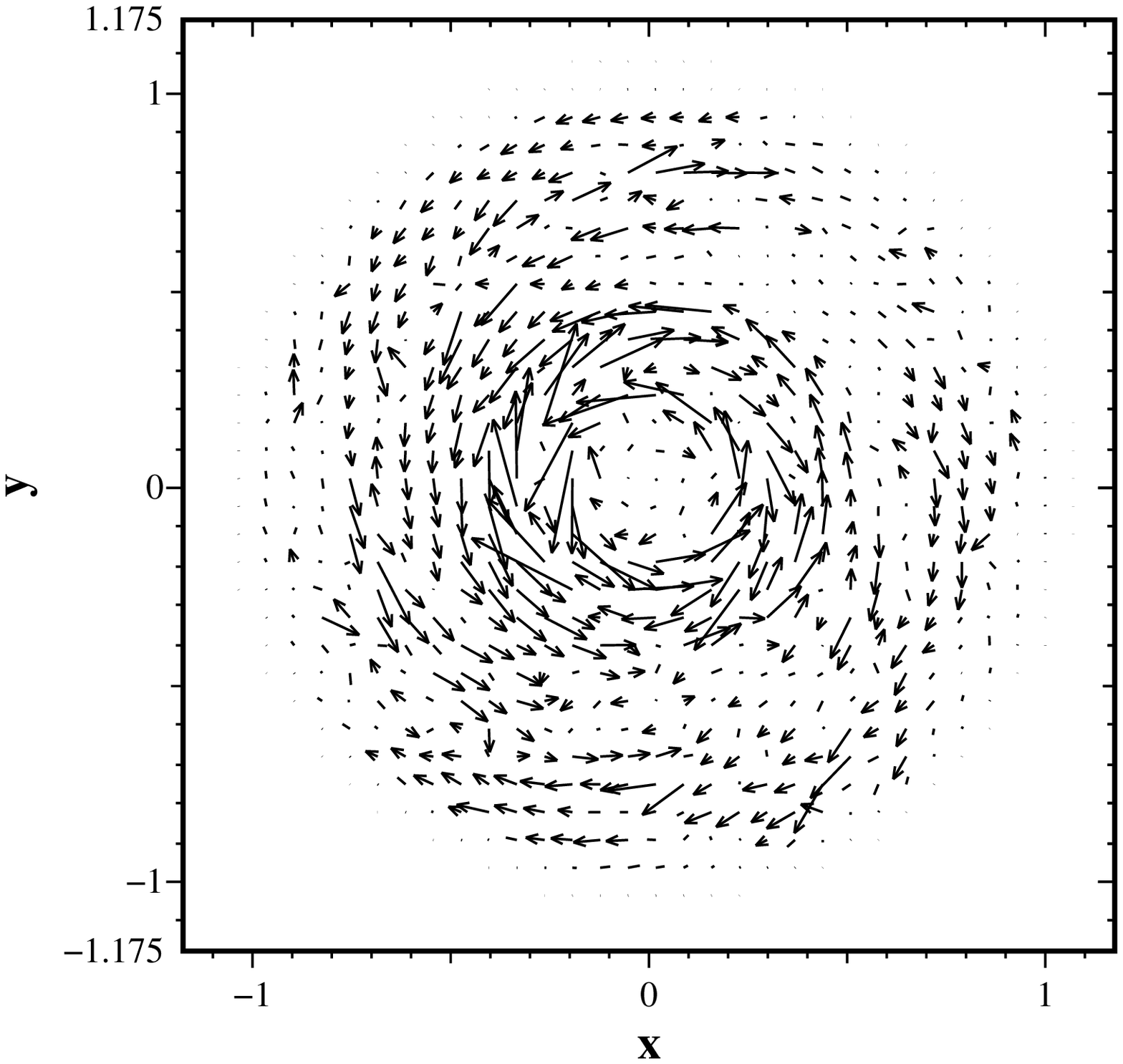} &
(b)\includegraphics[width=0.44\textwidth]{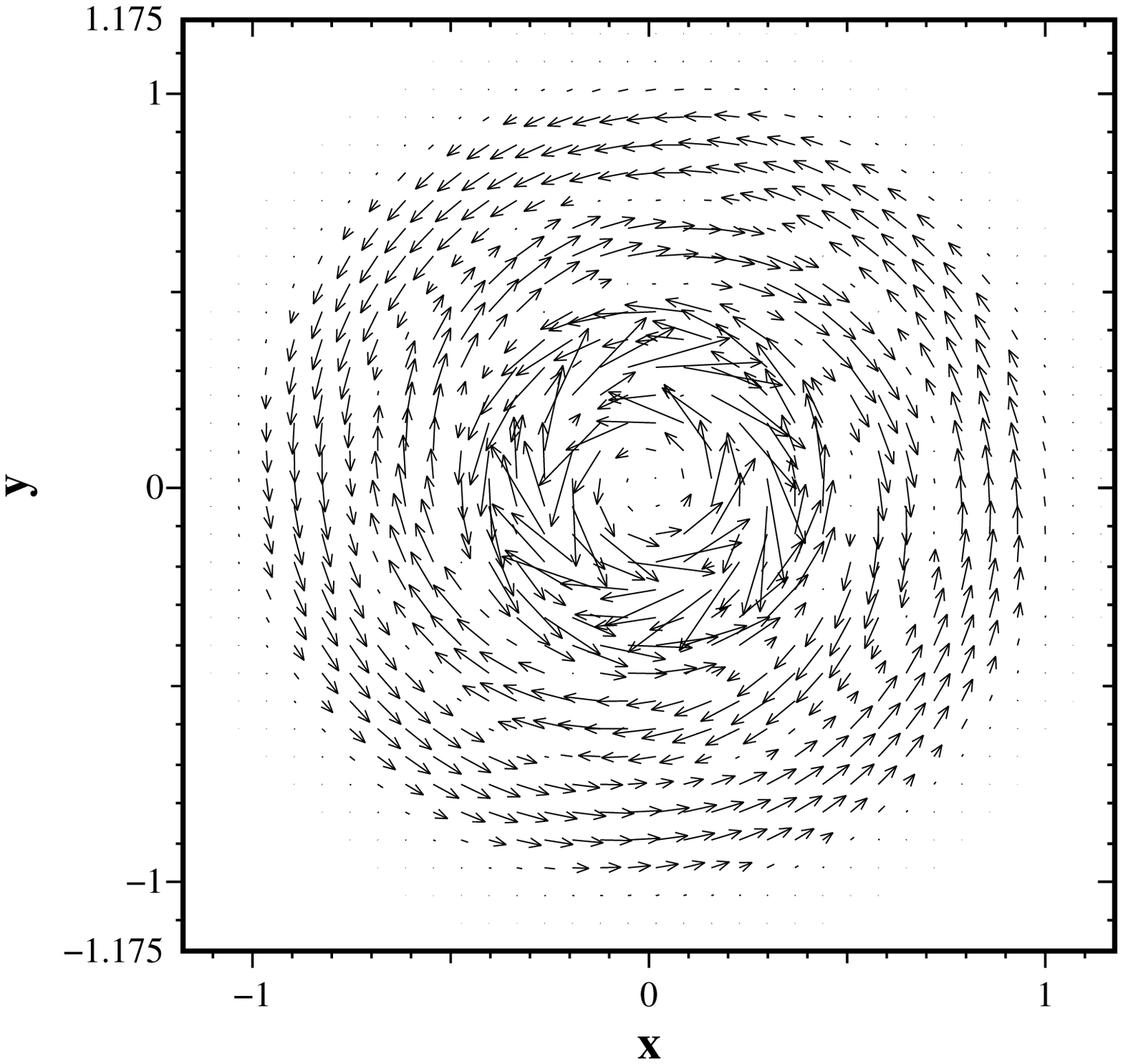} \\
\end{tabular}
\end{center}
\caption{(a) Statistically steady state of model~135 of
\protect{\cite{metal12}} with $R_\alpha=1, R_\omega=20$ at dimensionless time $t=17$ (approximately $13$ Gyr); (b) steady
state of similar model without the ongoing injections of random field
(NOINJECT10) at the same time.}
\label{fig1}
\end{figure*}

\begin{figure*}
\begin{center}
\vspace{0.5cm}
\begin{tabular}{lll}
(a)\includegraphics[width=0.30\textwidth]{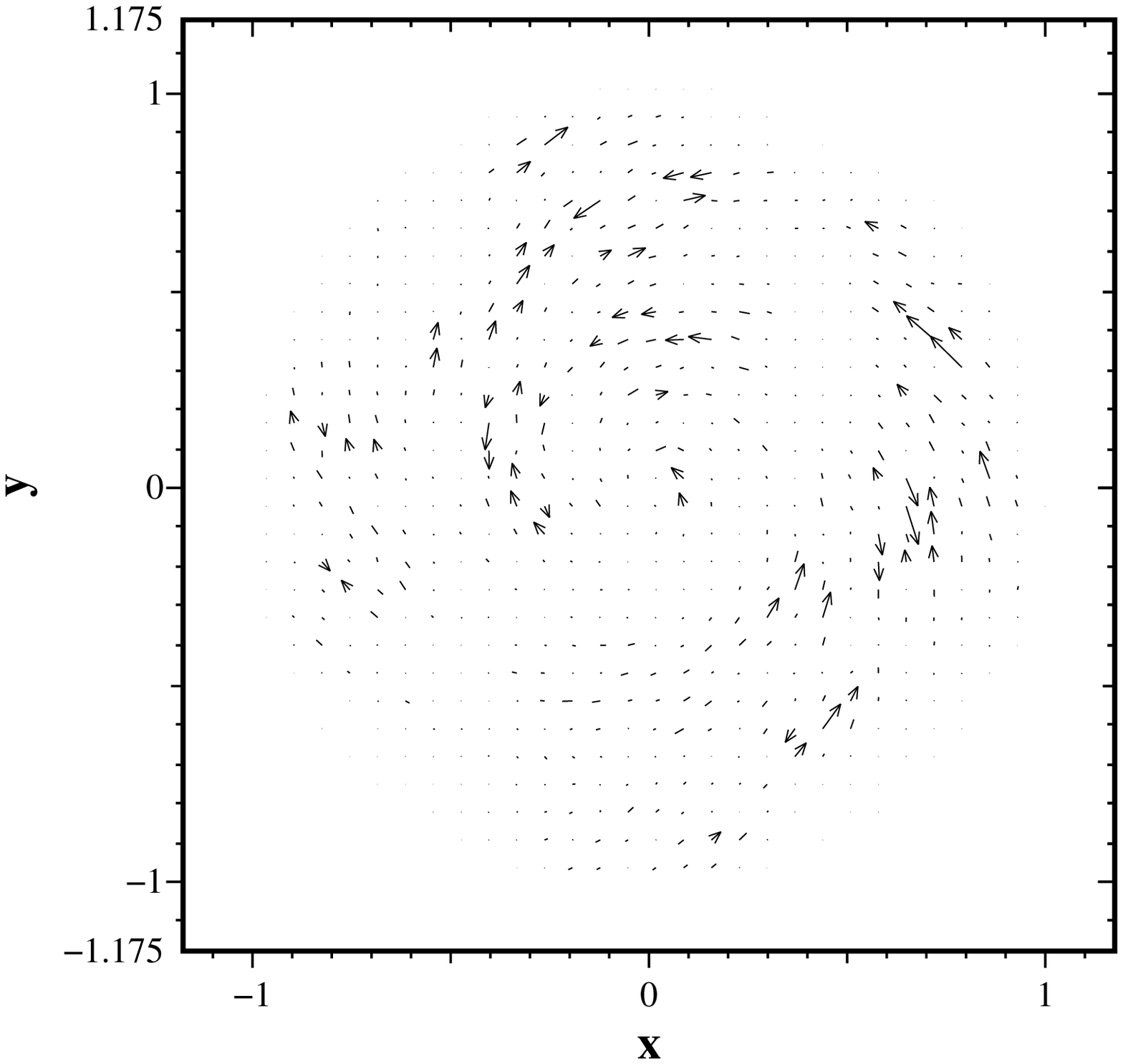} &
(b)\includegraphics[width=0.30\textwidth]{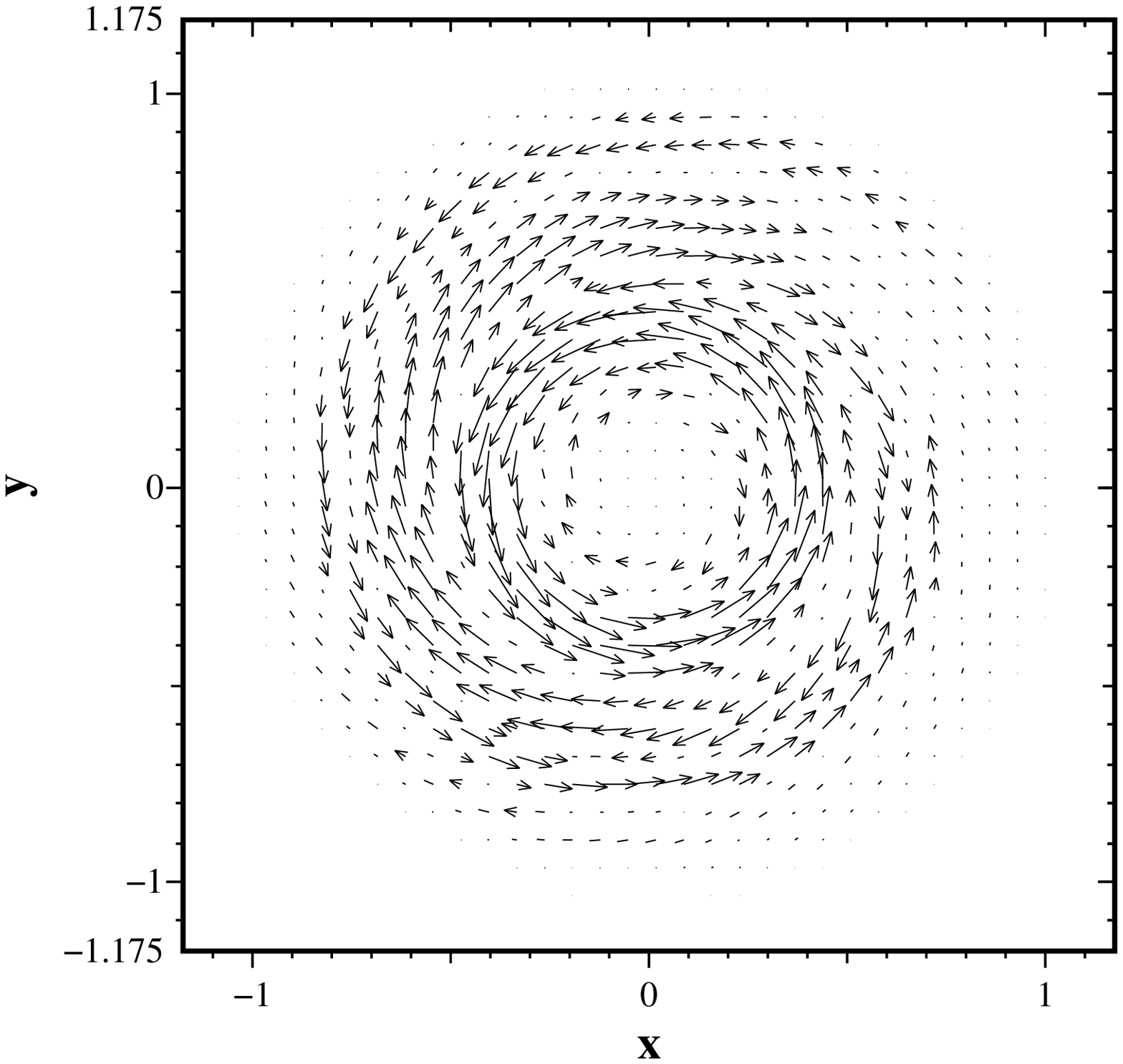} &
(c)\includegraphics[width=0.30\textwidth]{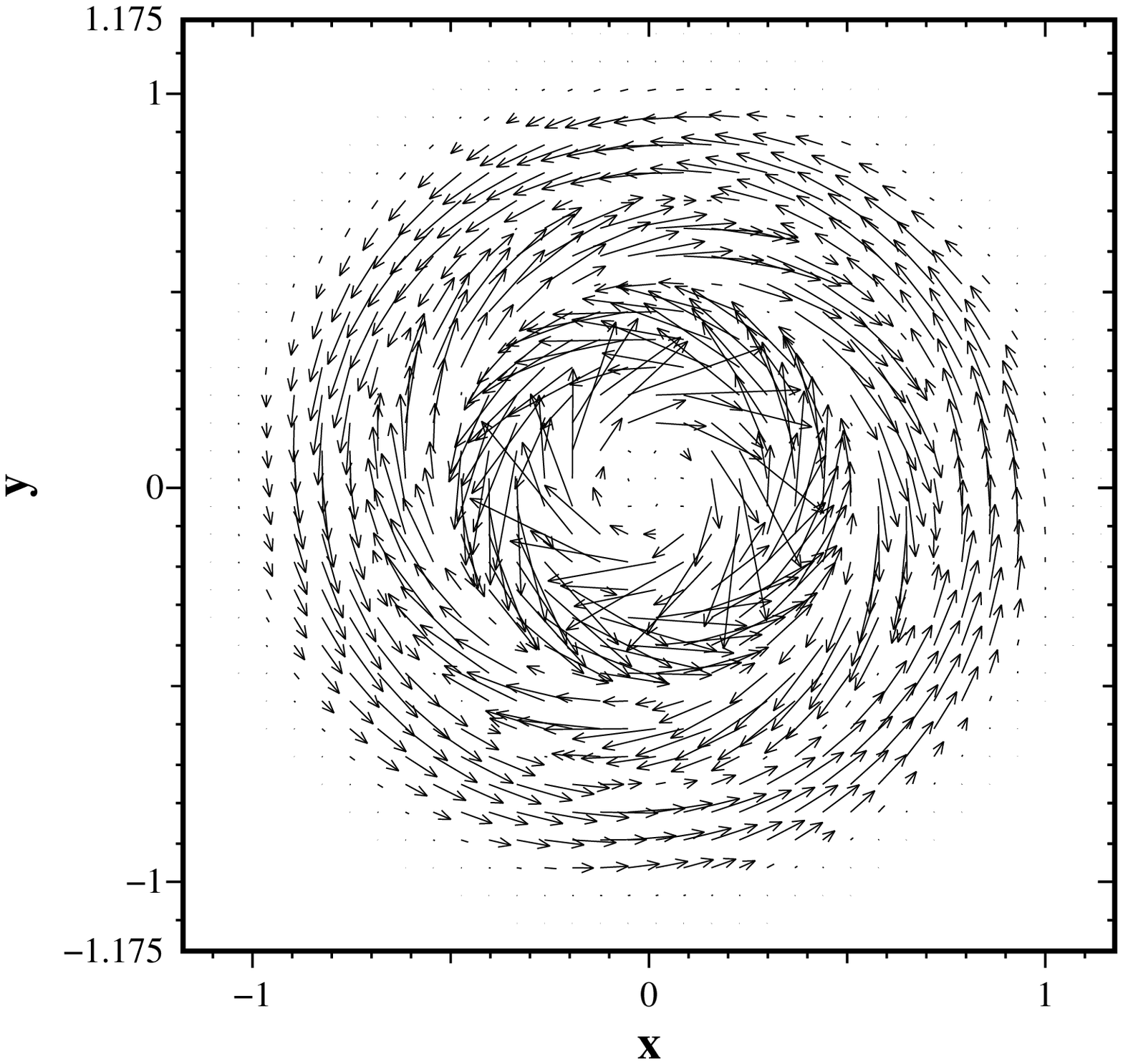} \\
\end{tabular}
\end{center}
\caption{Snapshots of the field configuration of the model NOINJECT10 at
times (a) $t=0.1$ (ca. $0.08$ Gyr);
(b) $t=1.0$ (ca. $0.8$ Gyr);
(c) $t=3.0$ (ca. $2.4$ Gyr).}
\label{timeseq}
\end{figure*}

We show in figure~\ref{fig1}a the statistically steady field (at age
ca. 13 Gyr) resulting from a standard computation of \cite{metal12}
(model~135, $R_\alpha=1, R_\omega=20$; we keep here and below the
numbers of the models from the working journal to allow
identification of the models between various papers). In this
case the
initial field is randomly distributed in $n_{\rm spot}=100$ discrete
regions, with r.m.s. value $B_0$ of the field in these regions of
order unity - i.e. the alpha-quenching nonlinearity is
immediately important.
Figure~\ref{initfield} shows a representation of
this initial field.
Large-scale reversals are clearly visible in
the resulting statistically steady configuration of this
model -- figure~\ref{fig1}a. We then show
in figure~\ref{fig1}b (model NOINJECT10) the results of a similar
computation using the same initial conditions, the only difference
being that there is now no ongoing injection of fields in discrete
regions. The final, steady, field is now smooth (the irregularities
visible in figure~\ref{fig1}a are caused by the ongoing injection of
small-scale field), but two large-scale reversals are again clearly
visible.
A reversal has appeared by dimensionless time $t=1$ (age $\lta 1$ Gyr),
and have stabilized by $t=3$ (ca $2.5$ Gyr). A time sequence is shown in figure~\ref{timeseq} for the model whose steady configuration is shown in
figure~\ref{fig1}b.

\begin{figure*}
\begin{center}
\vspace{0.5cm}
\begin{tabular}{ll}
(a)\includegraphics[width=0.44\textwidth]{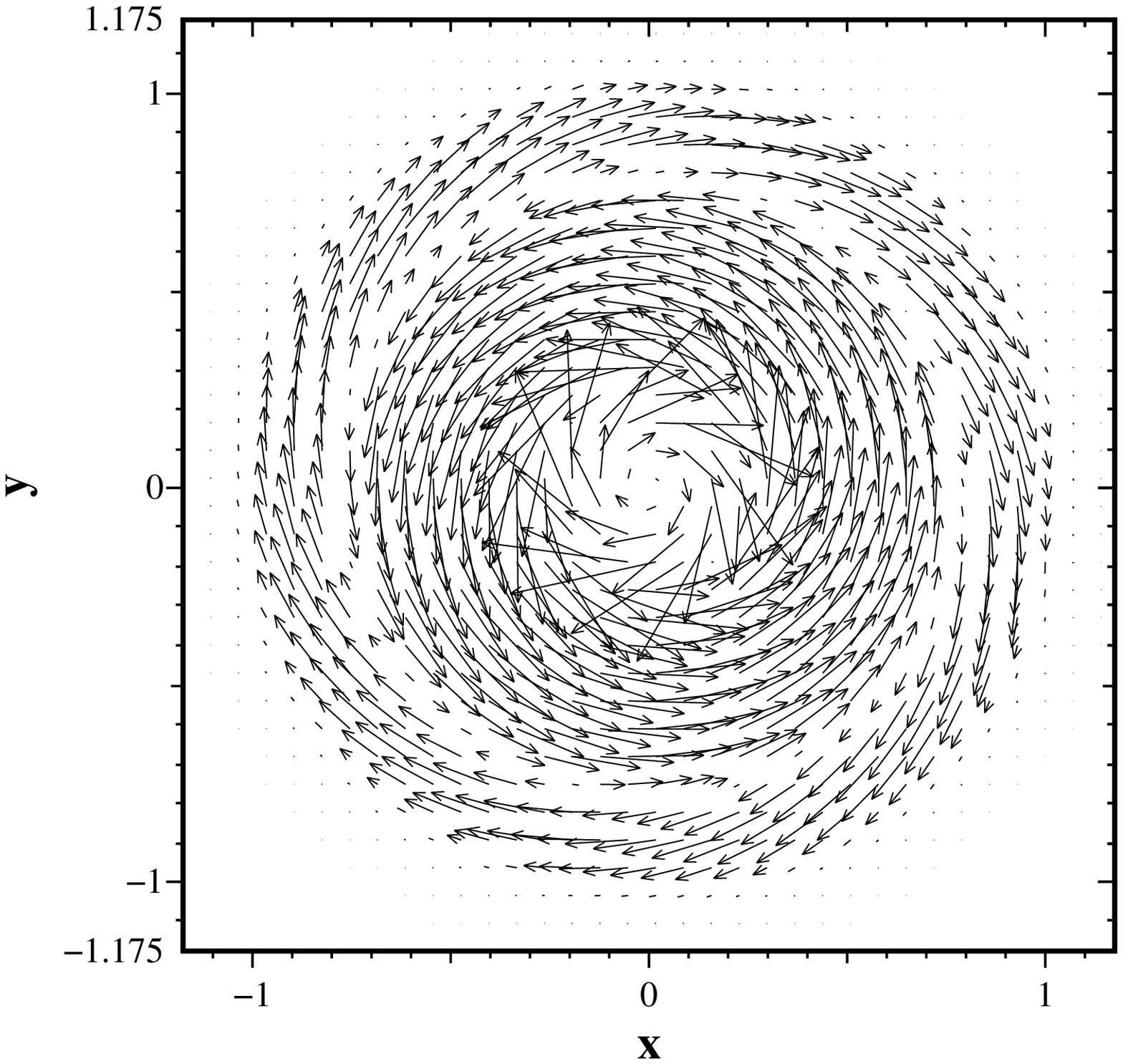} &
(b)\includegraphics[width=0.44\textwidth]{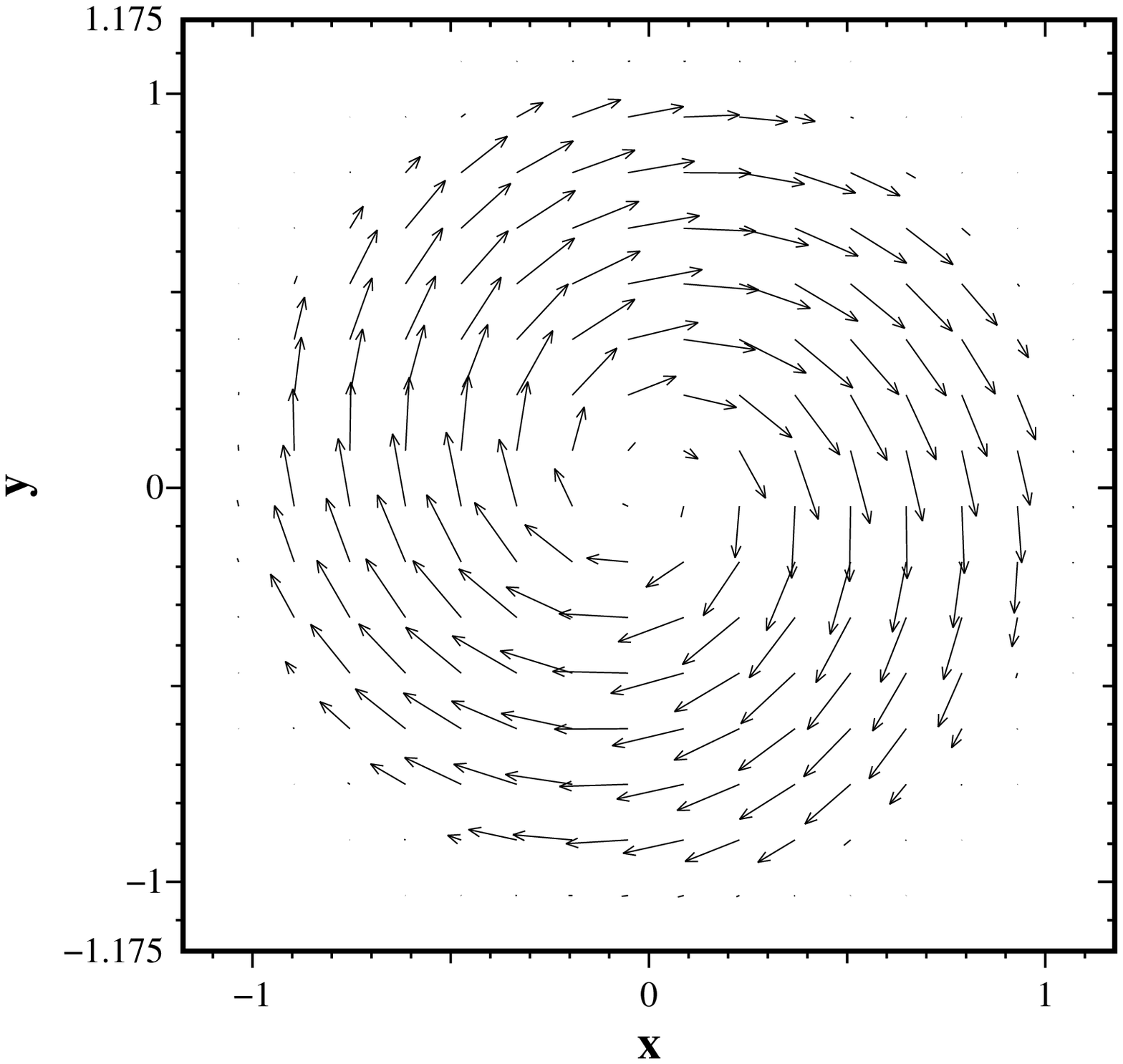} \cr
(c)\includegraphics[width=0.44\textwidth]{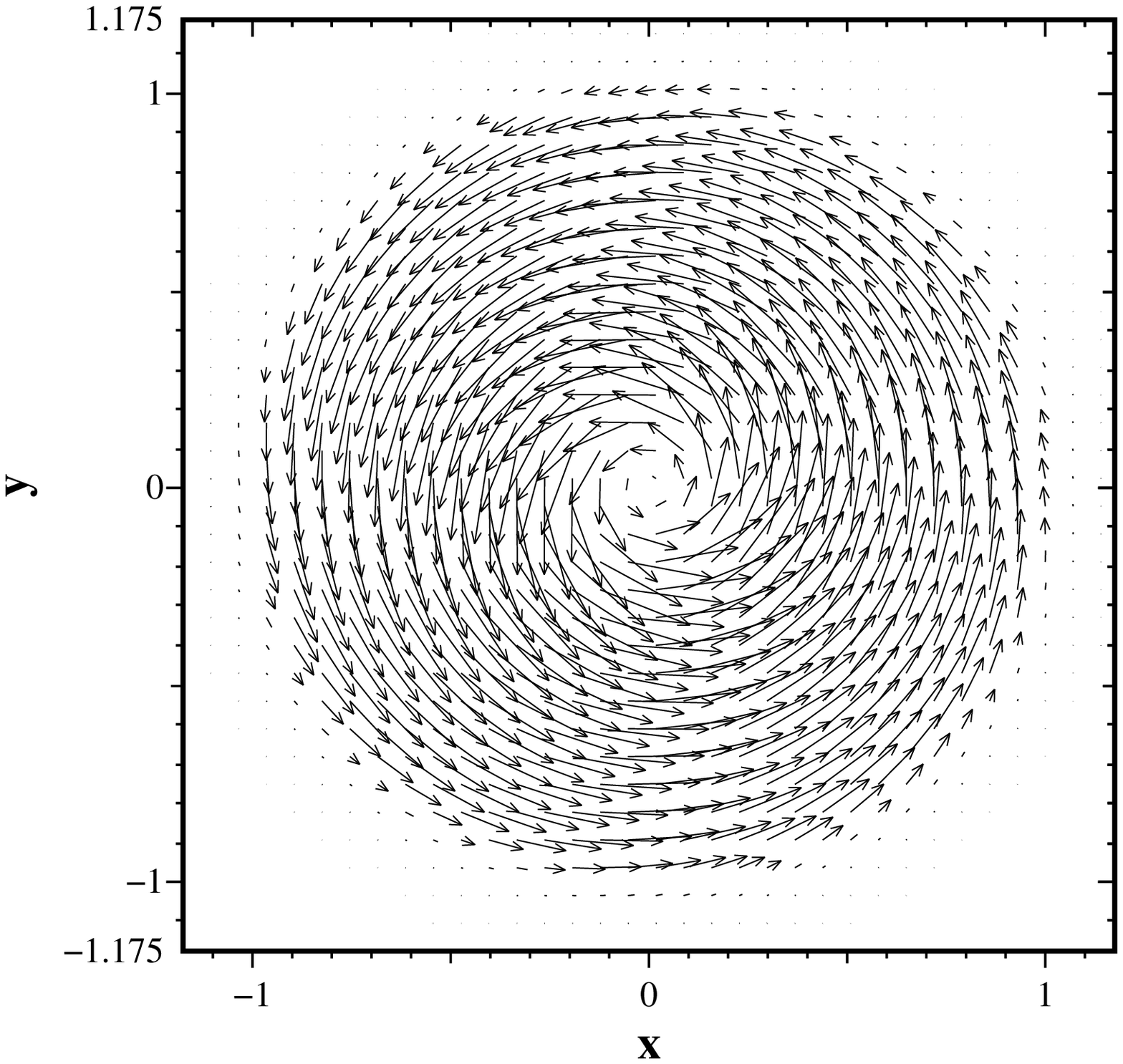} &
 (d)\includegraphics[width=0.44\textwidth]{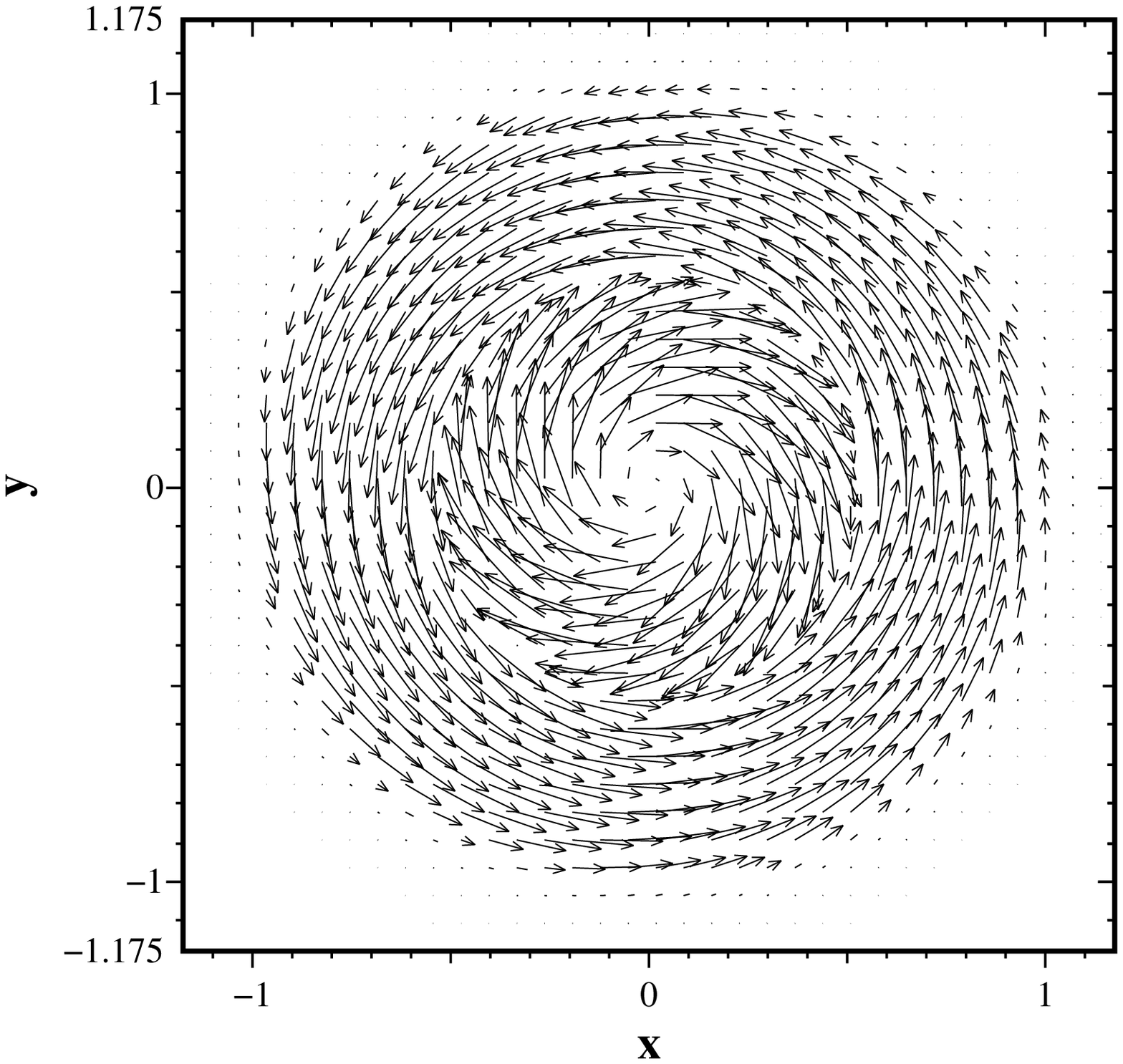}
\\
\end{tabular}
\end{center}
\caption{Steady field configurations for (a) $R_\alpha=1,
R_\omega=20$, r.m.s.~value of initial spotty field $B_0\approx
10^{-6}$; (b) $R_\alpha=1, R_\omega=20$, initial field randomly
assigned at every mesh point with r.m.s.~value of unity; (c) as (a)
with $B_0 \approx 10^{-8}$; (d) the model NOINJECT10 with a
different realization of random numbers -- compare figure~\ref{fig1}b.}
\label{fig2}
\end{figure*}

We expect that by reducing the strength of the seed field we will reduce the number
of reversals to zero. In order to test this prediction we
repeated the computation of figure~\ref{fig1}b with the r.m.s. value, $B_0$, of
the initial field in the discrete spots that was smaller by a factor
of $10^{-6}$, and then by a factor of $10^{-8}$. In the first of
these cases the steady configuration  has a single reversal
(figure~\ref{fig2}a, model NOINJECT11), in the second case there are
no reversals (figure~\ref{fig2}c, model NOINJECT16).

We also anticipate that a reduction of efficiency of the
large-scale dynamo will affect reversals in a more or less similar way
as reducing the seed field strength. The point is that both
of these changes act to prolong the stage of quasi-kinematic magnetic field
evolution. We verified this expectation, repeating  the above
computations  with $R_\omega$ reduced from $20$ to $10$; then with
$B_0\approx 1$ there is one reversal in the final steady
configuration,  and with $R_\omega=7$ there are no reversals.
With $B_0\approx 10^{-6}, R_\omega=20$ there are none.

If the size of the spots that localize the seed field
 becomes very small, reversals
disappear. We demonstrate this in an experiment where the initial field is
randomly assigned at {\it every} mesh point, again with $R_\alpha=1,
R_\omega=20$, and $B_0\approx 1$. The final steady field
configuration is shown in figure~\ref{fig2}b (model NOINJECT12) -- it
is smooth with no reversals. A similar result was found with
$B_0 \approx 10^{-6}$. Note that, in order to simplify comparison between
these models, we used the same realization of random numbers to determine the
seed field distribution in each case discussed above, with the exception of
the case shown in figure~\ref{fig2}b.

We further show that the eventual steady state configuration has some dependence
on the initial conditions, by the following experiment. The case NOINJECT10
(the same initial conditions as model 135 from \cite{metal12},
without the ongoing field injections) is re-simulated with a different
realization of the random numbers,  arriving at a
field configuration that is quite different from that shown in
figure~\ref{fig1}b -- see figure~\ref{fig2}d.
We also repeated the NOINJECT10 calculation with `global'
 alpha-quenching, i.e. quenching by the mean global energy.
The final configuration was quite different, but reversals are again present.
We conclude that the form of the initial field can influence significantly the steady state
configuration.

Repeating the case illustrated in figure~2b with $n_{\rm spot}=50$
produces a superficially quite similar final state.
As a final check, a simulation with $n_{\rm spot}=1, B_0=1$ produces a field without
reversals.

\section{Discussion and Conclusions}
\label{disc}

We deduce from these experiments that long-lived reversals can be
generated by a standard 2D (nonaxisymmetric) mean-field dynamo model,
given suitable
initial conditions and a suitable range of the dynamo governing
parameters. The initial conditions play a key role -- this is
probably why such solutions have not been widely noted previously.
Different steady solutions can be found for the same dynamo
parameters, depending on the initial conditions.
Large-scale reversals can appear after about $1$ Gyr, and they stabilize after $2-3$ Gyr.
The important
elements appear to be an inhomogeneous distribution of the initial
field and strong differential rotation (larger $R_\omega$), the
latter effect being enhanced by initial fields that are locally near
equipartition values. Note that even with larger values of $R_\omega$ ($R_\omega=20$ say),
some initial conditions still produce final configurations
without reversals.

We note that some of these results were anticipated by
\cite{becketal94}. This paper studied a one-dimensional model and
showed that, with a sufficiently strong small-scale seed field,
reversals could persist for times of order the age of the Universe.

As a straightforward consequence of the above conclusion, a
systematic observational search for magnetic field reversals in a
representative sample of spiral galaxies may provide valuable
information concerning seed magnetic fields, and so
clarify initial stages of galactic magnetic field evolution.
This appears to be a realistic possibility for
the forthcoming generation of radio telescopes.
For a contemporary comparison, it might be appropriate for orientation to look at figure 11 of
\cite{vaneck11}, showing their best model of the reversals
in the Milky Way field.

If we compare figure~\ref{fig1}a with subsequent
plots, the ongoing magnetic field injections gives an important
additional mechanism
to generate magnetic reversals. In addition to the global scale
magnetic field reversals discussed above, this model
contains several local reversals which appear to be due to local
field injections, and which exist for some time. Possibly, such local reversals
can be compared with reversals suggested to exist in the Milky Way 
\citep[see e.g.][]{han}

We appreciate that the models presented are very oversimplified,
but our intention is to show that initial conditions can strongly influence the
long term evolution of the large-scale magnetic field. With "standard"
weak initial fields ($|{\bm B}| << B_{\rm eq}$), any reversals are transient
features. It appears important that the nonlinearity operates {\it before}
the large-scale field forms.
 Of course, confirmation or revision of the conclusions of this paper in
the framework of 3D mean-field models, or even by direct numerical simulations
of the microscopic induction equation, performed with realistic
hydrodynamical models of very early galaxies would be an important
development of the theory of galactic dynamos. However this obviously is
beyond the scope of this paper.

We are grateful to our colleagues, who work on magnetic field
evolution in very young galaxies namely T.~Arshakian, R.~Beck, M.~Krause,
R.~Stepanov, for fruitful discussions which stimulated the writing of
this the paper.
We also thank Anvar Shukurov
and an anonymous referee for critical readings of the paper.

\end{document}